# The Correlative Method of Unsymmetrized Self-Consistent Field (CUSF)

# O Método Correlativo Não Simetrizado do Campo Autoconsistente (CUSF)

*Clóves Gonçalves Rodrigues, André Luiz Cardoso da Silva*
*Escola de Ciências Exatas e da Computação,*
*Pontifícia Universidade Católica de Goiás*

In this work we describe the Correlative Method of Unsymmetrized Self-Consistent Field (CUSF). This method is based on a set of nonlinear integrodifferential equations for the one-particle configurational distribution functions and for the self-consistent potentials of the atoms. Here we present the fundamental concepts of the CUSF, the hypotheses of the method, the basic equations, the self-consistent potential, the thermodynamics of the anharmonic crystalline solids, and the quantum corrections in the quasi-classical approximation.

## A.1 Conceitos Fundamentais

Num cristal os átomos realizam vibrações próximas a seus sítios da rede

$$\mathbf{r} = \hat{A}\mathbf{n} + \mathbf{q}, \tag{A.1}$$

sendo $\mathbf{r}$ a posição do átomo, $\hat{A}$ a matriz da rede, $\mathbf{n}$ vetores de componentes inteiros, e $\mathbf{q}$ o deslocamento dos átomos da sua posição de equilíbrio. As amplitudes efetivas são pequenas em comparação com a distância interatômica, ou seja, o parâmetro de Lindemann $\delta$ é sempre

$$\delta = \frac{1}{a}\sqrt{\langle q^2 \rangle} \ll 1 \tag{A.2}$$

onde $a$ é a distância interatômica média entre primeiros vizinhos. Na maioria dos cristais $\delta$ não excede 0.1 até a temperatura de fusão e unicamente para os assim chamados cristais quânticos (He, $H_2$) ele é por volta de 0.3 devido às grandes amplitudes de vibração de ponto zero. Portanto, a energia potencial pode ser expandida em uma série de potências dos deslocamentos atômicos dos sítios da rede

$$U(\mathbf{r}) = U_0 + \sum_{\ell \geq 2} U_\ell, \tag{A.3}$$

onde $U_0 = U(\hat{A}\mathbf{n})$, e

$$U_\ell = \frac{1}{\ell!} \sum_{\substack{\mathbf{n}_1 \cdots \mathbf{n}_\ell \\ 1 \leq \alpha_1, \cdots \alpha_\ell \leq 3}} U_{\alpha_1, \cdots \alpha_\ell}^{\mathbf{n}_1 \cdots \mathbf{n}_\ell} q_{\alpha_1}^{\mathbf{n}_1} \cdots q_{\alpha_\ell}^{\mathbf{n}_\ell}. \tag{A.4}$$

Os coeficientes desta série são as derivadas da energia potencial em relação às coordenadas atômicas quando todos os átomos estão em seus sítios da rede, e são conhecidos por coeficientes de força

$$U^{\mathbf{n}_1\cdots\mathbf{n}_\ell}_{\alpha_1,\cdots\alpha_\ell} = \left.\frac{\partial^\ell U}{\partial x^{\mathbf{n}_1}_{\alpha_1}\cdots\partial x^{\mathbf{n}_\ell}_{\alpha_\ell}}\right|_{|\mathbf{r}|=|\hat{A}\mathbf{n}|} . \qquad (A.5)$$

Para um cristal perfeito os termos lineares são nulos, ou seja, $U_1 = 0$. Por outro lado na presença de defeitos como vacâncias, superfícies, etc., esses coeficientes não são nulos e introduzem uma relaxação na rede aumentando as amplitudes de vibração na direção da anisotropia [1].

Como a série expressa em (A.3) converge ela pode ser aproximada por uma soma parcial

$$U(\{\mathbf{r}^{\bar{n}}\}) = U_0 + \sum_{\ell\geq 2}^{s} U_\ell . \qquad (A.6)$$

Os métodos da teoria estatística das propriedades atômicas dos cristais são definidos pelo valor de $s$ e pelos coeficientes de força dados em (A.5) [2]. Assim

$$\begin{cases} s=2, U^{\mathbf{mn}}_{\alpha\beta} = cte & \Rightarrow \text{aproximação harmônica} \\ s=2, U^{\mathbf{mn}}_{\alpha\beta} = f(a) & \Rightarrow \text{aproximação quase-harmônica} \\ s=4, |U_\ell| \ll U_2, \ell \geq 3 & \Rightarrow \text{aproximação de fraca anarmonicidade} \\ s=4, |U_3|, U_4 \sim U_2, \ell \geq 3 & \Rightarrow \text{aproximação de forte anarmonicidade} \end{cases}$$

O cálculo exato da integral estatística de um cristal pode ser feito somente nas aproximações harmônica e quase harmônica quando temos $s=2$. Geralmente, em baixas temperaturas os efeitos anarmônicos são considerados através de correções por teoria de perturbação nas aproximações citadas [3]. Isto conduz a expansões assintóticas a baixas temperaturas das funções termodinâmicas do cristal. Em temperaturas elevadas os efeitos anarmônicos nos cristais são fortes e tais aproximações se tornam inadequadas.

## A.2 Equações Básicas

As equações do método CUSF [4] são obtidas diretamente da equação de Liouville da Física Estatística Clássica, a qual dá a equação de evolução da função de distribuição do sistema $W(\mathbf{r}_1,\mathbf{p}_1,\ldots,\mathbf{r}_N,\mathbf{p}_N,t)$ que contém, em princípio, todas as informações físicas relevantes do sistema. Esta equação é dada por

$$\frac{\partial W}{\partial t} = \{H, W^{(N)}\}, \tag{A.7}$$

onde $(N)$ indica $N$ partículas e $\{H, W^{(N)}\}$ é o parêntese de Poisson da função de Hamilton $H$ com $W$, expresso por

$$\{H, W^{(N)}\} = \sum_{i=1}^{N}\left[\left(\nabla_{\mathbf{r}_i}H\right)\bullet\left(\nabla_{\mathbf{p}_i}W^{(N)}\right) - \left(\nabla_{\mathbf{p}_i}H\right)\bullet\left(\nabla_{\mathbf{r}_i}W^{(N)}\right)\right]. \tag{A.8}$$

Considerando um sistema de $N$ partículas contidas num volume $V$ que interagem entre si aos pares e por meio de forças centrais a função de Hamilton é dada por

$$H = \sum_{i=1}^{N}\frac{p_i^2}{2m} + U, \tag{A.9}$$

onde o primeiro termo do lado direito é a energia cinética total dos átomos e o segundo termo é a energia potencial de interação entre as partículas dado por

$$U = \frac{1}{2}\sum_{i\neq j}^{N}\Phi_{ij} = \frac{1}{2}\sum_{i=1}^{N}\sum_{j=1}^{N}(1-\delta_{ij})\Phi_{ij}(\mathbf{r}_i - \mathbf{r}_j), \tag{A.10}$$

onde $\delta_{ij}$ é igual a 1 para $i=j$ e zero para $i\neq j$ e $\Phi_{ij}$ representa a energia de interação entre dois pares de átomos na rede cristalina. Usando as equações de (A.1) até (A.10) podemos encontrar, após algumas manipulações, a seguinte expressão:

$$\frac{\partial W_i}{\partial t} + \frac{\mathbf{p}_i}{m}\bullet\left(\nabla_{\mathbf{r}_i}W_i\right) - \sum_{j=1}^{N}(1-\delta_{ij})\int\left(\nabla_{\mathbf{r}_i}\Phi_{ij}\right)\bullet\left(\nabla_{\mathbf{p}_i}W_i\right)d\mathbf{r}_j d\mathbf{p}_j = 0. \tag{A.11}$$

## A.3 Hipóteses do Método

As equações básicas do CUSF são obtidas a partir da equação (A.11) fazendo-se as seguintes hipóteses:

### # 1

A densidade de probabilidade do espaço de fase não é simétrica em relação às transposições das coordenadas e dos impulsos das partículas, ou seja:

$$W(\mathbf{r}_i,\mathbf{p}_i,\ldots,\mathbf{r}_j,\mathbf{p}_j) \neq W(\mathbf{r}_j,\mathbf{p}_j,\ldots,\mathbf{r}_i,\mathbf{p}_i). \tag{A.12}$$

### # 2

A função para duas partículas será o produto das funções para uma partícula

$$W(\mathbf{r}_i, \mathbf{p}_i, \mathbf{r}_j, \mathbf{p}_j; t) = W(\mathbf{r}_i, \mathbf{p}_i; t) \cdot W(\mathbf{r}_j, \mathbf{p}_j; t).  \tag{A.13}$$

Esta igualdade é conhecida como "*Condição de Multiplicidade*". Tal condição, porém, só é justificada quando diferentes funções de uma partícula não se interferem, ou seja, tal aproximação exclui as correlações dinâmicas. Para corrigir esta aproximação é utilizada a teoria de perturbação. Esta correção é que dá origem ao nome "*Correlativo*" no método.

### **# 3**

Considerando-se um estado de equilíbrio temos:

$$\frac{\partial W}{\partial t} = 0.  \tag{A.14}$$

Esta hipótese é válida se a transferência de partículas de uma célula para outra for desprezada durante o tempo de observação, isto é, estamos supondo que o sistema está em equilíbrio termodinâmico.

### **# 4**

Para um cristal perfeito quando $N \to \infty$ toda função de uma partícula pode ser considerada como tendo a mesma forma, mas com alguns deslocamentos regulares, isto é

$$W_k(\mathbf{r}_k, \mathbf{p}_k) = W_k(\mathbf{r}_k - \mathbf{a}_k, \mathbf{p}_k),  \tag{A.15}$$

onde $\mathbf{a}_k$ são as coordenadas dos sítios da rede, dependentes dos tipos de rede e dos seus parâmetros que serão determinados como consequência da teoria proposta. Esta hipótese significa que cada partícula da rede perfeita tem a mesma probabilidade espacial de ser encontrada com deslocamentos $\mathbf{r}_k - \mathbf{a}_k$ em relação ao ponto $\mathbf{a}_k$ da rede. Usando o método de separação de variáveis, onde admitimos que $W(\mathbf{q}, \mathbf{p}) = f(\mathbf{p}) \cdot w(\mathbf{q})$, e aplicando as hipóteses do método ao conjunto de equações (A.11) a equação (A.13) pode ser reescrita como

$$W(\mathbf{q}_i, \mathbf{p}_i) = (2\pi m\Theta)^{-3/2} e^{-p^2/2m\Theta} w(\mathbf{q}),  \tag{A.16}$$

onde $\Theta = k_B T$, sendo $k_B$ a constante de Boltzmann, $T$ a temperatura absoluta e $w(\mathbf{q})$ uma função que satisfaz a equação integral não-linear

$$\ln\{\lambda w(\mathbf{q})\} + \frac{1}{\Theta} \int \mathrm{K}(\mathbf{q} - \mathbf{q}') w(\mathbf{q}') d\mathbf{q}' = 0,  \tag{A.17}$$

onde $\lambda$ é uma constante de normalização determinada por

$$\int_{-\infty}^{\infty} w(\mathbf{q}) d\mathbf{q} = 1,  \tag{A.18}$$

e

$$\mathrm{K}(\mathbf{q} - \mathbf{q}') = \sum_{i,j=1}^{N} (1 - \delta_{ij}) \Phi(|\mathbf{a}_i + \mathbf{q} - (\mathbf{a}_j + \mathbf{q}')|),  \tag{A.19}$$

onde $\mathbf{q}$ e $\mathbf{q}'$ representam os deslocamentos das partículas em relação aos seus sítios da rede $\mathbf{a}_i$ e $\mathbf{a}_j$ respectivamente. As equações (A.16) e (A.17) são as equações fundamentais do CUSF.

## A.4 O Potencial Autoconsistente

O potencial autoconsistente de um átomo individual é [5]

$$u(\mathbf{q}) = \int K(\mathbf{q}-\mathbf{q}')w(\mathbf{q}')d\mathbf{q}' - \frac{1}{2}\int K(\mathbf{q}-\mathbf{q}')w(\mathbf{q})w(\mathbf{q}')d\mathbf{q}d\mathbf{q}', \qquad (A.20)$$

sendo, então, a energia potencial autoconsistente do cristal dada por

$$U^0(\mathbf{r}_1,\mathbf{r}_2,\ldots,\mathbf{r}_N) = \sum_{i=1}^{N} u_i(\mathbf{r}_i). \qquad (A.21)$$

Visto que no estado de equilíbrio termodinâmico os desvios dos átomos de seus sítios da rede são pequenos até a temperatura de fusão, a energia potencial do cristal, veja Eq. (A.3), pode ser expandida em série de Taylor em potências dos deslocamentos atômicos. Após fazer isto, o potencial autoconsistente de um átomo, veja Eq. (A.20), toma a forma:

$$u(\mathbf{q}) = u_0 + \sum_{\ell,m,n=0}^{\infty} \frac{1}{\ell!m!n!} F_{x^\ell y^m z^n} q_x^\ell q_y^m q_z^n, \qquad (A.22)$$

onde

$$F_{x^\ell y^m z^n} = u_0 + \sum_{i,j,k=0}^{\infty} \frac{(-1)^{i+j+k}}{i!j!k!} K_{x^{i+\ell} y^{j+m} z^{k+n}} \overline{q_x^i q_y^j q_z^k}, \qquad (A.23)$$

e

$$\overline{q_x^i q_y^j q_z^k} = \int q_x^i q_y^j q_z^k w(\mathbf{q})d\mathbf{q}, \qquad (A.24)$$

$$K_{x^\lambda y^\mu z^\nu} = \left.\frac{\partial^{\lambda+\mu+\nu}}{\partial q_{x^\lambda}\partial q_{y^\mu}\partial q_{z^\nu}} K(\mathbf{q})\right|_{\mathbf{q}=0}. \qquad (A.25)$$

As fórmulas de (A.22) até (A.24) constituem um grupo infinito de equações transcendentais para os momentos $\overline{q_x^i q_y^j q_z^k}$ da distribuição $w(\mathbf{q})$. Em virtude da convergência das séries de (A.22) e (A.23) elas podem ser aproximadas pela soma parcial

$$u(\mathbf{q}) = u_0 + \sum_{2\leq \ell+m+n\leq s} \frac{1}{\ell!m!n!} F_{x^\ell y^m z^n} q_x^\ell q_y^m q_z^n, \qquad (A.26)$$

e

$$F_{x^\ell y^m z^n} = u_0 + \sum_{i+j+k\leq s-\ell-m-n} \frac{(-1)^{i+j+k}}{i!j!k!} K_{x^{i+\ell} y^{j+m} z^{k+n}} \overline{q_x^i q_y^j q_z^k}. \qquad (A.27)$$

Assim, neste caso, a equação integral não linear (A.17) para $w(\mathbf{q})$ é reduzida a um grupo finito de equações transcendentais aproximadas (A.24), (A.26) e (A.27) para $\overline{q_x^i q_y^j q_z^k}$, com $i+j+k<s$, que podem ser resolvidas numericamente. Quando $s=2$ temos a aproximação harmônica e quase-harmônica, válidas para temperaturas muito baixas (com exceção dos cristais quânticos). Quando $s\geq 4$ [6] os termos anarmônicos são levados em consideração e a

---

solução destas equações transcendentais permite incluir a forte anarmonicidade na aproximação de ordem zero. Este caso é muito importante, pois, os termos anarmônicos de ordem superior são pequenos para alguns cristais até a temperatura de fusão [7].

## A.5 Termodinâmica dos Sólidos Cristalinos Anarmônicos

A termodinâmica dos cristais anarmônicos é formulada a partir da expressão

$$u_i(\mathbf{r}_i) = \tilde{u}_i(\mathbf{r}_i) - \frac{1}{2}\overline{\tilde{u}_i(\mathbf{r}_i)} \; , \tag{A.28}$$

onde,

$$\tilde{u}_i(\mathbf{r}_i) = \sum_j (1-\delta_{ij})\int \Phi^{ij}(|\mathbf{r}_i - \mathbf{r}_j|)w_i(\mathbf{r}_j)d\mathbf{r}_i \; , \tag{A.29}$$

$$\Lambda_i = \int e^{-\tilde{u}_i(\mathbf{r}_i)/\Theta}d\mathbf{r}_i \; , \tag{A.30}$$

$$w_i(\mathbf{r}_i) = \Lambda_i^{-1} e^{-\tilde{u}_i(\mathbf{r}_i)/\Theta} \; . \tag{A.31}$$

A equação (A.29) é o potencial médio autoconsistente de um átomo devido a todos os outros em seu entorno. Este potencial tem sua forma obtida a partir do "*princípio variacional de Bogoliubov*" [8]. A soma de todos os potenciais autoconsistentes individuais

$$U^0(\mathbf{r}_1,\mathbf{r}_2,...,\mathbf{r}_N) = \sum_{i=1}^N u_i(\mathbf{r}_i), \tag{A.32}$$

é chamado de "*energia potencial autoconsistente*". Diferentemente da energia potencial exata (A.10), ela depende da temperatura em virtude das equações (A.28) a (A.31) dependerem da temperatura, isto é, $\partial U^0/\partial\Theta \neq 0$. Nesta aproximação o valor médio da energia autoconsistente é igual ao valor médio da energia exata:

$$\overline{U^0} = \frac{1}{2}\sum_{i\neq j}\int \Phi^{ij} w_i(\mathbf{r}_i)w_j(\mathbf{r}_j)\,d\mathbf{r}_i d\mathbf{r}_j = \overline{\mathbf{U}} \; . \tag{A.33}$$

Já foi provado, que no caso de uma repulsão forte a pequenas distâncias interatômicas, a função de distribuição de uma partícula é diferente de zero somente naquela região onde as funções de todos os outros átomos são iguais a zero e por esta razão as integrais (A.33) sempre convergem. Além disso, quando a temperatura tende a zero, as funções de uma partícula passam a serem fortemente localizadas em torno dos sítios de equilíbrio da rede e, consequentemente a energia autoconsistente tende a energia exata do cristal.

Devemos mostrar que este método garante a compatibilidade termodinâmica dos resultados. Visto que a energia autoconsistente (A.32) é dependente da temperatura, não é evidente que a função

---

[7] G. Leibfried, W. Ludwig, *Theory of anharmonic effects in crystals*, Academic Press, New York, USA, 1961; G. Leibfried, *Gittertheorie der mechanischen und thermischen eingenschaften der kristalle*, Springer-Verlag, Berlin, Germany, 1955.

[8] N. N. Bogoliubov, *Kinetic equations*, Journal of Experimental and Theoretical Physics, v. 8, n. 16, pp. 691-702, 1946; N. N. Bogoliubov, *Kinetic equations*, Journal of Physics USSR, 3, 10, 265-274, 1946 (in Russian); N. N. Bogoliubov, K. P. Gurov, *Kinetic equations in quantum mechanics*, Journal of Experimental and Theoretical Physics, v. 7, n. 17, pp. 614-628, 1947 (in Russian).

$$F^0 = -\Theta \ln \int \exp\left[-\frac{1}{\Theta}\sum_{i=1}^{N}\frac{p_i^2}{2m_i} + U^0\right]\frac{d\mathbf{r}_1\mathbf{p}_1\ldots d\mathbf{r}_N d\mathbf{p}_N}{(2\pi\hbar)^{Nd}} \qquad (A.34)$$

é a "energia de livre de Helmholtz". Como se sabe, a energia livre deve satisfazer as equações de estado calorífico e térmico. Na equação (A.34) o termo $(2\pi\hbar)^{Nd}$ é a função de partição semi-clássica e $d$ é a dimensão do sólido cristalino. Substituindo (A.34) na equação de "Gibbs-Helmholtz"

$$-\Theta^2\frac{\partial}{\partial\Theta}\left(\frac{F^0}{\Theta}\right) = \frac{3}{2}N\Theta + \overline{U^0} - \Theta\overline{\frac{\partial U^0}{\partial \Theta}} \qquad (A.35)$$

verifica-se a existência de um termo adicional, $-\Theta\,\overline{\partial U^0/\partial\Theta}$, que em princípio deveria violar a equação de Gibbs-Helmholtz. Derivando (A.34) em relação à temperatura $\Theta$, e em seguida calculando sua média, é possível mostrar que este termo é nulo, e portanto a energia livre (A.34) satisfaz a equação de Gibbs-Helmholtz. Derivando (A.34) em relação à coordenada termodinâmica $a_\mu$

$$\frac{\partial F^0}{\partial a_\mu} = \overline{\frac{\partial U^0}{\partial a_\mu}} \qquad (A.36)$$

obtemos uma relação entre a derivada da energia livre e a média da derivada da energia autoconsistente. Em seguida, calculando a derivada no segundo membro

$$\frac{\partial U^0}{\partial a_\mu} = \sum_{i\neq j}\frac{\partial u_i}{\partial a_\mu} =$$
$$= \sum_{i\neq j}\left\{\int\left(\frac{\partial \Phi^{ij}}{\partial a_\mu}w_j + \Phi^{ij}\frac{\partial w_j}{\partial a_\mu}\right)d\mathbf{r}_j - \frac{1}{2}\int \Phi^{ij}\left[\frac{\partial \Phi^{ij}}{\partial a_\mu}w_i w_j + \Phi^{ij}\left(\frac{\partial \Phi^{ij}}{\partial a_\mu}w_j + \Phi^{ij}\frac{\partial w_j}{\partial a_\mu}\right)\right]d\mathbf{r}_j d\mathbf{r}_i\right\} \qquad (A.37)$$

e sua média será

$$\overline{\frac{\partial U^0}{\partial a_\mu}} = \overline{\frac{\partial U}{\partial a_\mu}} \ . \qquad (A.38)$$

Substituindo (A.38) em (A.36), verificamos que a equação (A.34) também satisfaz a "equação de estado térmico". Portanto, a energia livre na aproximação zero do CUSF tem a mesma forma usual da energia de Gibbs e a correção à aproximação zero deve ser feita pelo método usual da teoria de perturbações. Além disso, como ela foi definida é possível mostrar, utilizando o princípio variacional estatístico de Bogoliubov, que a energia autoconsistente dá a melhor estimativa para a energia livre, $F^0 \geq F$, entre todos os valores aproximados, calculados por meio da soma dos potenciais de uma partícula. Ou seja, a aproximação do CUSF minimiza o erro, que está sempre presente quando se utiliza métodos aproximados.

As funções termodinâmicas podem ser expressas em termos dos potenciais autoconsistentes (A.29): a energia livre de Helmholtz

$$F^0 = -\Theta\sum_i \ln\left\{\left(\frac{m\Theta}{2\pi\hbar^2}\right)^{3/2}\int \exp\left[-\frac{u_i(\mathbf{r}_i)}{\Theta}\right]d\mathbf{r}_i\right\}, \qquad (A.39)$$

a energia

$$E^0 = -\Theta^2 \frac{\partial}{\partial \Theta}\left(\frac{F^0}{\Theta}\right) = \sum_i \left\{\frac{3\Theta}{2} + u_i(\mathbf{r}_i)\right\}, \quad (A.40)$$

e a entropia,

$$S^0 = -k_B \ln \overline{W_0^{(N)}} = -k_B \overline{\ln \prod_{i=1}^{N} W_i(\mathbf{r}_i, \mathbf{p}_i)}. \quad (A.41)$$

Logo, a entropia também é proporcional ao valor estatístico do logarítmo da densidade de probabilidade.

No caso de sólidos com imperfeições (superfícies, vacâncias, etc.) é utilizada a ideia original de Gibbs e as funções termodinâmicas das imperfeições são tratadas com quantidades excedentes. Dois sistemas para os sólidos são considerados, um com as camadas de interface reais e um outro que é chamado de "sistema de referência", nos quais as duas fases permanecem homogêneas até a superfície de separação. Escolhendo uma superfície de separação equimolecuar, a energia livre de um sistema componente com interface plana é dada por,

$$F_s^0 = F_1^0 + F_2^0 + \sigma^0 A, \quad (A.42)$$

onde $F_i^0$ são as energia livres das fases homogênea nos sistemas de referência, $\sigma$ é a densidade de energia livre da superfície e $A$ é a área da interface. Assim as propriedades termodinâmicas de superfícies e vacâncias podem ser calculadas como, por exemplo, a densidade de energia de superfície e a tensão superficial.

Segundo a terminologia de Gibbs, as funções termodinâmicas de um sólido cristalino imperfeito são definidas como sendo a diferença entre as funções de um cristal imperfeito e um cristal perfeito (sem o defeito), ambos constituídos por um mesmo número de átomos. Nesta definição, o sistema de comparação é o cristal perfeito e as propriedades são funções de excesso. Portanto, a energia livre de formação de um defeito é

$$F_0^f = \Theta \sum_i Z_i \ln\left[\frac{\int e^{-u(\mathbf{r})/\Theta} d\mathbf{r}}{\int e^{-u_i(\mathbf{r}_i)/\Theta} d\mathbf{r}_i}\right], \quad (A.45)$$

onde $Z_i$ é número de coordenação.

## A.6 Correções Quânticas: Aproximação Quase-Clássica

Os métodos clássicos da teoria estatística de sólidos anarmônicos são válidos para temperaturas acima da temperatura de Debye, $T_D$. Em Temperaturas mais baixas é necessário levar em conta correções quânticas. No intervalo, $T_D/3 < T < T_D$, estes efeitos são pequenos e podem ser tratados como correções. Neste caso, as correções quânticas às equações clássicas expressam-se através de potências pares da constante de Planck. Temos exemplos disso, nos trabalhos feitos por Wigner [9], Uhlenbeck e Gropper [10] e Kirkwood [11] em que uma

---

[9] E. Wigner, *On the quantum correction for thermodynamic equilibrium*, Phys. Rev., v. 40, p. 749, 1932.
[10] G. E. Uhlenbeck, L. Gropper, *The equation of state of a non-ideal Einstein-Bose or Fermi-Dirac gas*, Phys. Rev.,

expansão de ondas planas é usada para estimar, aproximadamente, a função de partição canônica, resultando em uma expressão para a energia livre em uma série de potências de $\hbar$. Na teoria do CUSF existe um formalismo parecido [12]. Nesta teoria cada átomo do sólido é descrito por uma "*matriz densidade*" que é não-simétrica com respeito à troca de coordenadas entre partículas idênticas,

$$\hat{\varrho}(...,j,...,k,...) \neq \hat{\varrho}(...,k,...,j,...) , \tag{A.46}$$

valendo a condição de multiplicidade

$$\hat{\varrho}(1,...,N) = \prod_{i=1}^{N} \hat{\varrho}_i(\mathbf{r}_i) . \tag{A.47}$$

Na aproximação quântica do CUSF, o hamiltoniano exato $\hat{H}$ de um sistema de $N$ partículas é substituído pelo hamiltoniano autoconsistente, ou seja

$$\hat{H} \longrightarrow \hat{H}^0 = \sum_{i=1}^{N} \left[ -\frac{\hbar^2}{2m}\nabla_i^2 + u_i^Q(\mathbf{r}_i) \right] , \tag{A.48}$$

que depende de um potencial autoconsistente quântico $u^Q(\mathbf{r}_i)$ e possui algumas peculiaridades. Em primeiro lugar, ele depende da temperatura, pois, os potenciais autoconsistentes contêm $\Theta$. Além disso, o hamiltoniano não comuta com sua derivada em relação à temperatura,

$$\left[ \frac{\partial \hat{H}^0}{\partial \Theta}, \hat{H}^0 \right] \neq 0, \tag{A.49}$$

e seus dois termos dependem da constante de Planck. A função de partição autoconsistente tem a forma,

$$Z = \prod_{i=1}^{N} \frac{1}{(2\pi\hbar)^3} \int \exp\left(-\frac{i}{\hbar}\mathbf{r}_i \bullet \mathbf{p}_i\right) \Omega_i(\mathbf{r}_i, \mathbf{p}_i, \Theta, \alpha) d\mathbf{r}_i d\mathbf{p}_i , \tag{A.50}$$

onde

$$\Omega_i(\mathbf{r}_i, \mathbf{p}_i, \Theta, \alpha) = \exp\left(-\frac{\alpha}{\Theta}\hat{H}_i\right) \exp\left(\frac{i}{\hbar}\mathbf{r}_i \bullet \mathbf{p}_i\right) . \tag{A.51}$$

Derivando $\Omega_i$ com respeito a $\alpha$ obtem-se a equação

$$\frac{\partial \Omega_i}{\partial \alpha} = -\hat{H}_i \Omega . \tag{A.52}$$

Esta equação se diferencia da forma de Block por um fator $1/\Theta$. A solução de $\Omega_i$ com a condição de Cauchy $\Omega_i \big|_{\alpha=0} = \exp(\mathbf{r}_i \bullet \mathbf{p}_i / \hbar)$ e $\alpha = 1$ tem a forma,

---

$$\Omega_i \approx \exp\left\{-\frac{1}{\Theta}\left[\frac{p_i^2}{2m} + u_i^Q(\mathbf{r}_i)\right] + \frac{i}{\hbar}\mathbf{r}_i\cdot\mathbf{p}_i\right\}\left(1 + \sum_{n\geq 1}\hbar^n \chi_n^{(i)}\right), \quad (A.53)$$

onde as funções $\chi_i^{(n)}$ são expressas através dos potenciais autoconsistentes $u_i^Q(\mathbf{r}_i)$. Por causa da segunda peculiaridade, temos também que expandir os potenciais autoconsistentes em séries de potências de $\hbar$,

$$u_i(\mathbf{r}_i) \approx u_i(\mathbf{r}_i) + \sum_{n\geq 1}\hbar^{2n} u_{i,n}(\mathbf{r}_i), \quad (A.54)$$

em que $u_i(\mathbf{r}_i)$ são os potenciais autoconsistentes clássicos.

A transformada de Fourier da equação (A.50) determina os elementos diagonais da matriz de densidade não-normalizada de uma partícula até $\hbar^2$

$$\varrho_i(\mathbf{r}_i) \approx w_i(\mathbf{r}_i)\exp\left[1 - \frac{\hbar^2}{12m\Theta^2}\xi_i(\mathbf{r}_i)\right], \quad (A.55)$$

onde

$$w_k(\mathbf{r}_i) = \Lambda^{-1} e^{-u_i(\mathbf{r}_i)/\Theta} \quad (A.56)$$

são as distribuições de uma partícula clássicas e $\xi_i(\mathbf{r}_i)$ são funções que contêm as correções quânticas à matriz densidade em termos dos potenciais autoconsistentes clássicos $u_i(\mathbf{r}_i)$. As funções $\xi_i(\mathbf{r}_i)$ são determinadas a partir da condição de normalização da matriz densidade na forma,

$$\begin{aligned}\xi_i(\mathbf{r}_i) = \xi_i^0(\mathbf{r}_i) &+ \frac{1}{\Theta}\left[\overline{u_i(\mathbf{r}_i)\xi_i(\mathbf{r}_i)} - \overline{u_i(\mathbf{r}_i)}\overline{\xi_i(\mathbf{r}_i)}\right] + \frac{1}{\Theta}\sum_j \overline{\xi_j(\mathbf{r}_i)}u_{ij}(\mathbf{r}_i)\\ &-\frac{1}{\Theta}\sum_j(1-\delta_{ij})\left[\int \Phi^{ij}(|\mathbf{r}_i - \mathbf{r}_j|)w_j(\mathbf{r}_j)\xi_j(\mathbf{r}_j)d\mathbf{r}_j\right.\\ &\left.-\frac{1}{2}\int \Phi^{ij}(|\mathbf{r}_j - \mathbf{r}_i|)w_i(\mathbf{r}_i)w_j(\mathbf{r}_j)d\mathbf{r}_i d\mathbf{r}_j\right],\end{aligned}$$
(A.57)

onde,

$$u_{ij}(\mathbf{r}_i) = \int \Phi^{ij}(|\mathbf{r}_j - \mathbf{r}_i|)w_j(\mathbf{r}_j)d\mathbf{r}_j - \frac{1}{2}\int \Phi^{ij}(|\mathbf{r}_j - \mathbf{r}_i|)w_i(\mathbf{r}_i)w_j(\mathbf{r}_j)d\mathbf{r}_i d\mathbf{r}_j, \quad (A.58)$$

$$\xi_i^0(\mathbf{r}_i) = \nabla^2 u_k(\mathbf{r}_k) - \frac{1}{2\Theta}[u_k(\nabla\mathbf{r}_k)]^2, \quad (A.59)$$

$$u_i(\mathbf{r}_i) = \sum_j u_{ij}(\mathbf{r}_i). \quad (A.60)$$

As equações (A.57) são equações integrais de Fredholm de segunda espécie. Para solucionar estas equações devemos notar as seguintes propriedades de $\xi_i(\mathbf{r}_i)$:

$$\sum_i \overline{\xi_i(\mathbf{r}_i)} = \sum_i \overline{\xi_i^0(\mathbf{r}_i)} = \frac{1}{2}\sum_i \overline{\nabla^2 u_i(\mathbf{r}_i)}. \quad (A.61)$$

Destas equações, na aproximação do campo autoconsistente, obtêm-se a primeira correção quântica para a energia livre

$$F^Q = \frac{\hbar^2}{24m\Theta} \sum_i \overline{\nabla^2 u_i(\mathbf{r}_i)} \ , \tag{A.62}$$

e as médias estatística de funções $f(\mathbf{r}_i)$ são definidas pelas relações,

$$\langle f(\mathbf{r}_i) \rangle = \overline{f(\mathbf{r})}_i - \frac{\hbar^2}{12m\Theta}\left[\overline{f_i(\mathbf{r}_i)\xi_i(\mathbf{r}_i)} - \overline{f_i(\mathbf{r}_i)}\,\overline{\xi_i(\mathbf{r}_i)}\right] \ . \tag{A.63}$$

A equações (A.55) a (A.63) formam a aproximação quase-clássica do CUSF. Estas equações quando solucionadas permitem obter a primeira correção às propriedades dinâmicas e termodinâmicas de um sólido cristalino. É importante ressaltar, que além da aproximação quase-clássica no CUSF existe o formalismo quântico, com equações do tipo Hartree e obedece ao princípio variacional estatístico de Bogoliubov, sendo que a solução destas equações para um sólido cristalino é bem mais complicada que a aproximação quase-clássica.